\documentstyle[aps,prl,epsfig,multicol,times]{revtex}
\makeatletter
\newlength\abovecaptionskip \newlength\belowcaptionskip
\def\@makecaption#1#2{%
 \vskip\abovecaptionskip \sbox\@tempboxa{#1: #2}%
 \ifdim \wd\@tempboxa >\hsize #1: #2\par \else \global \@minipagefalse 
 \hb@xt@\hsize{\hfil\box\@tempboxa\hfil}%
 \fi \vskip\belowcaptionskip} \makeatother

\newcommand{\km}{k_{\rm min}}
\newcommand{\kma}{k_{\rm max}}

\newcommand{\fr}{\frac}

\begin{document}
\title{Excess Modes in the Vibrational Spectrum of Disordered 
Systems and the Boson Peak}
\author{Jan W.~Kantelhardt, Stefanie Russ, and Armin Bunde}
\address{Institut f\"ur Theoretische Physik III, 
Universit\"at Giessen, D-35392 Giessen, Germany}
\draft
\maketitle
\begin{multicols}{2}[%
\begin{abstract}
We study a disordered vibrational model system, where the spring constants 
$k$ are chosen from a distribution $P(k) \propto 1/k$ above a cut-off value
$k_{\rm{min}}>0$.  We can motivate this distribution by the presence of free
volume in glassy materials.  We show that the model system reproduces several 
important features of the boson peak in real glasses: (i) a low-frequency excess
contribution to the Debye density of states, (ii) the hump of the specific 
heat $c_V(T)$ including the power-law relation between height and position 
of the hump, and (iii) the transition to localized modes well above the 
boson peak frequency. 
\end{abstract}
\pacs{PACS numbers: 63.50.+x, 65.60.+a}]

The vibrational spectrum and the corresponding thermal properties
of a large number of disordered materials exhibit two
characteristic anomalies \cite{zeller,stephens}.  While the
anomalous linear low-temperature specific heat can be described
well by the model of two-level systems \cite{tls}, the origin of
the second important anomaly, the boson peak, is still under
discussion.  It is not clear, if both anomalies are related
\cite{sokolov97}.  The boson peak refers to an excess contribution
to the usual Debye density of states (DOS) at low frequencies 
\cite{buch84,buch86,sok92meyer,dixon1,sugai,sokolov95}.  In
silicate and germanate glasses, for example, the boson peak 
frequency $\omega_{\rm peak}$ was reported between 4.5 and 13.5
THz, about $16-40$ times below the upper band edge $\omega_{\rm
E}$ \cite{dixon1,sugai}.  In silicate glasses 
\cite{sokolov97,buch84,sokolov95} and polybutadiene and
polysterene glasses \cite{sokolov95}, the peaks occur at
frequencies about $4-9$ times smaller than the corresponding Debye
frequency $\omega_{\rm D}$.  

The boson peak also shows up as a small hump in the scaled 
specific heat $c_V(T)/T^3$ in the temperature range $2-30$ K 
\cite{zeller,stephens,sokolov97,buch84,buch86,liu96}.  
The humps for different materials have a common shape. 
Position and height follow a power-law dependence, 
\begin{equation} \label{liu_pow}
c_V(T_{\rm hump})/T^3_{\rm hump} \sim T_{\rm hump}^{-x}
\end{equation}
with $x$ close to $2$ \cite{liu96} (see Fig.~3(b)).  The hump in
the specific heat is more pronounced for strong and intermediate
than for fragile glasses \cite{sokolov97}.  Apart from molecular
dynamics simulations on relatively small systems that do not yet
give a clear picture of the origin of the boson peak \cite{md},
relaxational and vibrational models have been proposed.  Many
experiments indicate that the relaxation-type soft potential model
\cite{softlaird} is appropriate for fragile glasses, while
vibrational models are more appropriate for strong and
intermediate glasses \cite{zus1,sokolov93} (see also 
\cite{frangoetz2}).  

In this paper, we do not wish to enter the controversial discussion of the 
microscopic origin of the boson peak. Instead, we concentrate on an idealized 
disordered vibrational model system that can be solved by standard 
numerical procedures.
We show that the model system is able to reproduce several important 
features of the boson peak in glassy materials.

Our starting point is the same as the one of Schirmacher et al.  
\cite{schirm}, who considered a simple cubic lattice of coupled
harmonic oscillators with random scalar spring constants between
nearest-neighbor masses. The spring constants were chosen from a truncated
Gaussian distribution with a lower cut-off value $\km$.
As a result of the disorder, a hump in the scaled DOS 
$Z(\omega)/\omega^2$ occurs at relatively high frequencies.
When shifting $\km$ towards negative values (which
imposes some unrealistic instability to all infinite lattices)
the hump is shifted towards frequencies about $2-3$ times smaller than 
$\omega_{\rm D}$, which is still far from the frequency range where the 
boson peak occurs.
Model systems with homogeneous distributions of positive spring
constants (see below) yield maxima in the same frequency range as
the Gaussian model system.  

In this work we also consider a disordered vibrational model system. 
The scalar spring constants $k$ are chosen from a power-law distribution 
$P(k) \propto 1/k$ above a positive cut-off value $\km$. 
We can motivate $P(k)$ by assuming a distribution of the free volume. 
We find that a low-frequency peak occurs in the DOS at $\omega_{\rm peak}$, 
in a frequency range comparable to the one of the boson peak of real glasses.
We expect that in our model system external pressure will shift the peak towards 
larger frequencies, an effect which has also been observed experimentally.
Investigating the localization properties of our model system we find that 
the modes around the peak are extended.  Localized modes occur above 
a crossover frequency $\omega_c \approx 3\omega_{\rm{peak}}$. 
This interesting property seems to be in line with experiments in real 
glasses.  The peak at $\omega_{\rm peak}$ is also visible in $c_V(T)$ as 
a small hump, which shows similar features as the boson peak in real 
systems, among them the power-law dependence (\ref{liu_pow}).

To be specific, we chose the spring constants from a normalized distribution
function
\begin{equation} \label{power}
P(k) = \fr{1}{\ln (\kma / \km)} \fr{1}{k} \qquad k \in [\km,\kma],
\end{equation}
where $\kma /\km$, the only parameter in the distribution, controls the amount 
of small spring constants. The form of Eq.~(\ref{power}) can be motivated as 
follows:  Consider a disordered system, where the neighbor distances 
$a_{ij}$ between the particles are in the interval $[a_{\rm min},a_{\rm max}]$.  
The limits $a_{\rm min}$ and $a_{\rm max}$ are related to
the size of the particles and the maximum diameter of holes in the
system, respectively, and thus characterize to some extent the distribution of 
the free volume in the system. For simplicity, we assume that the 
distances are distributed homogeneously between $a_{\rm min}$ and $a_{\rm max}$. 
When the distances $a_{ij}$ between two neighboring masses $i$ and $j$ 
fluctuate, the spring constants $k_{ij}$ between them also fluctuate. 
We assume that the $k_{ij}$ decay roughly exponentially with $a_{ij}$, $k_{ij} 
\approx \km \exp[(a_{\rm max}-a_{ij})/a^*]$ with a characteristic decay length
$a^*$, which might be a reasonable assumption in a strong or intermediate glass
with covalent binding \cite{sokolov93} and negligible charge separations.
Combining both assumptions, we arrive at Eq.~(\ref{power}), with the control 
parameter
\begin{equation} \label{ratio}
\kma /\km = \exp[(a_{\rm max}-a_{\rm min})/a^*].
\end{equation}
Note that the distribution (\ref{power}) is equivalent to
an homogeneous distribution of the logarithm of spring constants,
$P(\log k) = \rm{const}$. 

We have calculated numerically the DOS of this model system, where in a 
simple cubic lattice unit
masses $m$ at nearest-neighbor sites $i$ and $j$ are connected by
springs $k_{i j}$ chosen from $P(k)$.  Assuming scalar coupling
constants $k_{i j}$, the different components of the displacements
decouple and we obtain the same equations of motion 
\begin{equation} \label{motion} 
m \frac{d^2 u_j(t)}{dt^2} = \sum_i k_{ij} \left[u_i(t) - u_j(t) 
\right], 
\end{equation} 
for all spatial components of $u_j$.  The sum runs over the
nearest neighbor sites $i$ of site $j$.  The ansatz $u_j(t) = 
\psi^\alpha_j \exp(-i \omega_\alpha t)$ leads to an homogeneous 
system of equations for the $N$ unknown $\psi^\alpha_j$, from 
which the $N$ real eigenvalues $\omega_\alpha^2 \geq 0$ and the 
corresponding eigenvectors $(\psi_1^\alpha, \ldots,
\psi_N^\alpha)$, $\alpha = 1, \ldots, N$ can be determined.

\begin{figure}\centering 
\epsfxsize8.5cm\epsfbox{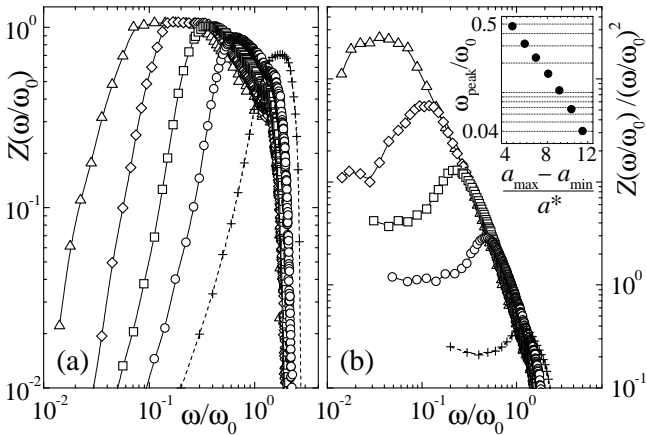} 
\parbox{8.5cm}{\caption[]{\small 
(a) Density of states (DOS) $Z(\omega/\omega_0)$ and (b) rescaled
DOS $Z(\omega/\omega_0)/(\omega/\omega_0)^2$ versus
$\omega/\omega_0$ for our model system ($\omega_0^2 \equiv
\kma/m$).  The symbols correspond to $\km /\kma = 10^{-5}$
($\bigtriangleup$), $10^{-4}$ ($\Diamond$), $10^{-3}$ ($\Box$),
and $10^{-2}$ ({\Large$\circ$}).  The numerical results for an
homogeneous distribution $k \in [0,\kma]$ are also shown (dashed
curves with $+$ symbols).  In the
inset, the frequency positions $\omega_{\rm peak}/\omega_0$ of
the maxima of $Z(\omega/\omega_0) /(\omega/\omega_0)^2$ are shown
versus the relative maximum size of holes $(a_{\rm max}-a_{\rm
min})/a^* = \ln (\kma /\km)$. } 
\label{fig:1}}\end{figure} 

To calculate the vibrational properties and the related specific
heat $c_V$ of the model system, we have employed the method of Williams
and Maris \cite{wm} and the Lanczos algorithm \cite{lanczos}.  In
Fig.~1 we show the DOS $Z(\omega/\omega_0)$, with $\omega_0 \equiv 
\sqrt{\kma/m}$, and the rescaled DOS $Z(\omega/\omega_0)/
(\omega/\omega_0)^2$ for several values of $\km/\kma$ between $10^{-2}$ 
and $10^{-5}$ \cite{footnote} and system sizes of up
to $65^3$ masses.  For all values of $\km/\kma$, both quantities
show a broad maximum at frequencies $\omega_{\rm peak}$, which
become smaller, if $\km /\kma$ is decreased.  Well below
$\omega_{\rm peak}$, we observe the conventional Debye behavior,
$Z(\omega) \propto \omega^2$.  For comparison we also calculated
the DOS of a vibrational model system with an homogeneous distribution of
spring constants $k \in [0,\kma]$ where the maximum in the DOS
simply results from a broadening of the van Hove singularity.  
The systems with power-law distributed 
$k_{ij}$ have their maxima in the low-frequency regime, where also the boson
peak in glasses is observed. Estimating the Debye frequency $\omega_D$ from
the plateaux in the scaled DOS (Fig.~1(b)), we find for the maxima in 
$Z(\omega)/ \omega^2$ the values 
$\omega_D/\omega_{\rm peak} \approx 3$ for $\km =10^{-2}$, 
$\omega_D/\omega_{\rm peak}\approx 4.5$ for $\km =10^{-3}$ and 
$\omega_D/\omega_{\rm peak}\approx 7$ for $\km =10^{-4}$ and still larger values
for smaller $\km$.  In contrast, the peak for the
homogeneous distribution, shows up at frequencies $\omega_{\rm peak}$
only a factor of $2$ below $\omega_{\rm D}$.  

When $\km/\kma$ is enhanced, the maximum in $Z(\omega)$ tends to
higher frequencies.  Since in our model system
$\km/\kma$ is related to the free volume parameter $(a_{\rm max}-a_{\rm min})$ 
via Eq.~(\ref{ratio}), the peak is shifted towards higher frequencies, when 
$(a_{\rm max}-a_{\rm min})$ is reduced.
We can imagine that $(a_{\rm max}-a_{\rm min})$ can be reduced by applying
external pressure to the system.
Indeed, experiments on glasses indicate that the boson peak is shifted to
higher frequencies, when the glasses were permanently desified
under pressure \cite{sugai}. This is in line with the results of our 
model system.

\begin{figure}\centering 
\epsfxsize8.5cm\epsfbox{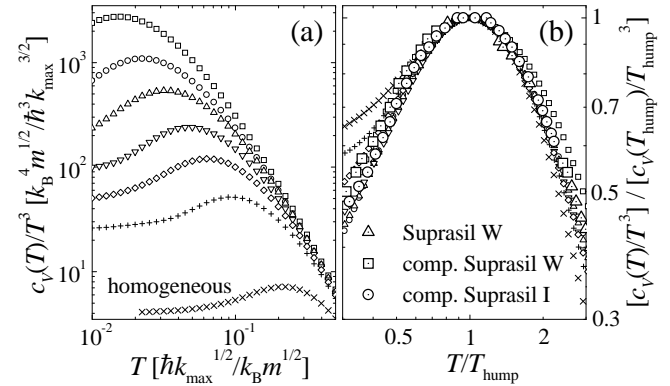} 
\parbox{8.5cm}{\caption[]{\small (a) Rescaled specific heat 
$c_V(T)/T^3$ versus temperature $T$ for $\km /\kma = 3 \times 
10^{-5}$ ($\Box$), $10^{-4}$ ({\Large $\circ$}), $3 \times
10^{-4}$ ($\bigtriangleup$), $10^{-3}$ ($\bigtriangledown$), $3
\times 10^{-3}$ ($\Diamond$), and $10^{-2}$ ($+$) and for an
homogeneous distribution $k \in [0,\kma]$ ($\times$).  In (b) the
results from (a) are rescaled by dividing $T$ by $T_{\rm hump}$
and $c_V(T)/T^3$ by $c_V(T_{\rm hump}) /T_{\rm hump}^3$.  
Deviations from the common shape occur for the smallest and the
largest parameter as well as for the homogeneous distribution of
spring constants.  The experimental values for several silica
glasses (large symbols, redrawn from Fig.~1 of \cite{liu96}), are
also shown. } 
\label{fig:2}}\end{figure} 

Next we consider the specific heat $c_V(T)$, which is related to
$Z(\omega)$ by:
\begin{equation} \label{cv}
c_V(T) = k_B \int_0^\infty d \omega \,
Z(\omega) \left(\frac{\hbar \omega}{k_B T}\right)^2
\frac{e^{\hbar \omega/k_B T}}{(e^{\hbar \omega/k_B T} -1)^2}.  
\end{equation}
 
Figure 2(a) shows $c_V(T)/T^3$ versus temperature $T$ for several
values of $\km /\kma$ as well as for the homogeneous distribution
of spring constants.  As expected from the behavior of $Z(\omega)$, 
a maximum in $c_V(T)/T^3$ occurs at $T_{\rm hump}$, which is shifted
towards lower temperatures and increases in height, when $\km
/\kma$ decreases.  For testing for a common shape of the hump, we
have plotted $[c_V(T)/T^3] /[c_V(T_{\rm hump}) /T_{\rm hump}^3]$
in Fig.~2(b) as a function of $T/T_{\rm hump}$ and compared the
results with the experimental curves for several vitreous silica
\cite{liu96}.  It is interesting that the results for our
model system at intermediate $\km /\kma$ values agree reasonably
well with the experimental data.  No fitting parameter was
involved.  The figure shows also that major deviations from the common 
shape occur for the homogeneous model system in the low-temperature range.  

\begin{figure}\centering 
\epsfxsize8.5cm\epsfbox{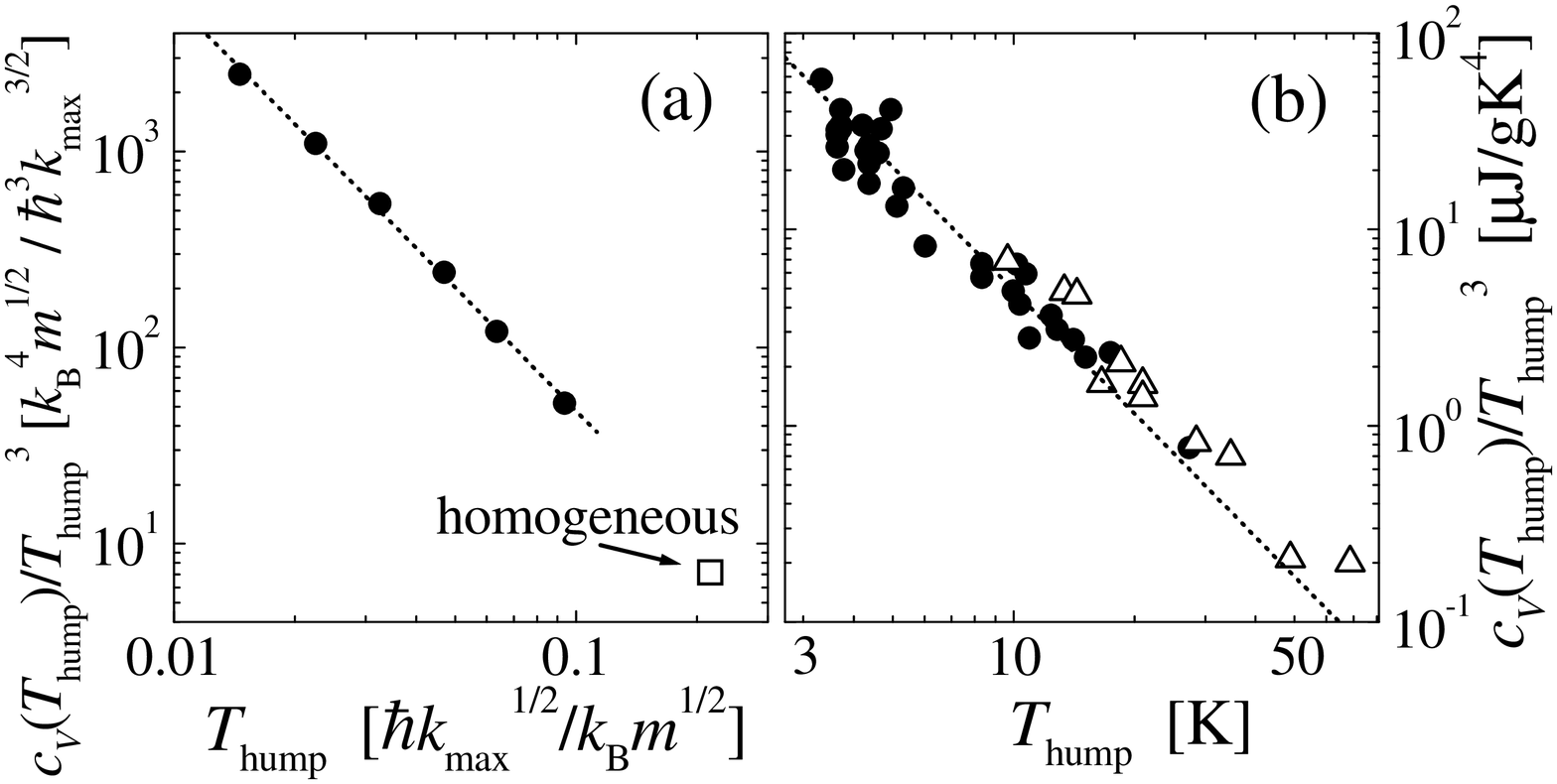} 
\parbox{8.5cm}{\caption[]{\small (a) The values of the maxima of
$c_V(T)/T^3$ from Fig.~2(a) versus the positions $T_{\rm hump}$ 
of the maxima.  In the log-log plot the data fall onto a straight 
line described by a power-law: $c_V(T_{\rm hump})/T^3_{\rm hump} 
\sim T_{\rm hump}^{-x}$ with $x \approx 2.1$ (dotted line).  (b)
Experimental results for several glasses and amorphous solids
({\Large$\bullet$}) as well as for some crystalline materials 
($\bigtriangleup$), redrawn after Fig.~2 of \cite{liu96}.  The
data for the glasses and the amorphous solids are consistent with
our theoretical results $x \approx 2.1$ (dotted line).  If also 
crystalline systems are included, one obtains a smaller
exponent.   Note that the scales of both axes are only changed by
constant  factors in (a) and (b), so the range is exactly the
same. } 
\label{fig:3}}\end{figure} 
 
To see, if the specific heat also satisfies the power-law relation 
(\ref{liu_pow}) we have plotted $c_V(T_{\rm
hump})/T^3_{\rm hump}$ as a function of $T_{\rm hump}$ in a
double-logarithmic fashion.  The result, shown in Fig.~3(a), 
agrees surprisingly well with the experimental data for a large 
number of glassy and amorphous solids, shown in Fig.~3(b).  
The basic power-law (\ref{liu_pow}) is reproduced with 
an exponent $x$ that is, within the error bars, identical with the 
one of real glasses.
Note  that both, Figs.~3(a) and 3(b), show the same range for the 
$x$- and $y$-axes.  Again, no fitting parameter was involved.  
It is interesting to note that also for crystalline systems 
the same power-law behavior has been found experimentally. 
An explanation of this is beyond the scope of this paper.

Finally we turn to the localization behavior of the vibrational
modes in our model system.  By general arguments, we can expect 
extended phonons for
small frequencies $\omega < \omega_{\rm c}$ and localized modes
for high frequencies $\omega > \omega_{\rm c}$.  The localization
mechanism for vibrations is quite similar to the Anderson
localization of quantum particles in disordered solids (see also
\cite{kantelhardt98,akita}) and it is interesting to see, if the 
modes in the peak regime fall into the extended or into the localized 
range.

To estimate the ratio between $\omega_{\rm peak}$ and 
$\omega_{\rm c}$, we have employed the method of
level statistics \cite{guhr}.  Level statistics (with a fixed
system size) have been applied earlier to the Gaussian model system
\cite{schirm}, indicating extended modes in the vicinity of 
$\omega_{\rm peak}$.  In the method, one first calculates the 
eigenvalues $\omega_\alpha^2$ of the vibrational equation and
then determines the eigenvalue spacings $s_\alpha =
(\omega_\alpha^2 - \omega_{\alpha-1}^2) /\Delta$, where $\Delta$
is the mean eigenvalue spacing in the frequency range considered. 
The dependence of the mean squared eigenvalue spacing $\langle s^2
\rangle$ on the system size $N$ indicates the localization
behavior.  With increasing system size, $J_0 \equiv {1\over
2}\langle s^2 \rangle$ tends to $J_0^{\rm Wigner} \approx 0.637$
for extended modes, while it approaches $J_0^{\rm Poisson} = 1$
for localized modes.  
Figure 4 shows the results of this analysis for $\km /\kma =
10^{-2}$ and $10^{-3}$.  From the figure we can deduce 
$\omega_{\rm c} \approx 1.3 \,\omega_0$ for $\km /\kma=10^{-2}$
and $\omega_{\rm c} \approx 0.7 \,\omega_0$ for $\km
/\kma=10^{-3}$.  For $\km /\kma=10^{-4}$ we obtain $\omega_{\rm c}
\approx 0.4 \,\omega_0$ in an analogous way.  A comparison with
the position of the peak (taken from the inset of Fig.~1(b))
yields $\omega_{\rm c} \approx 2.7 \,\omega_{\rm peak}$,
$\omega_{\rm c} \approx 3.1 \,\omega_{\rm peak}$, and $\omega_{\rm
c} \approx 3.9 \,\omega_{\rm peak}$ for $\km /\kma = 10^{-2}$,
$10^{-3}$, and $10^{-4}$, respectively.  

\begin{figure}\centering 
\epsfxsize8.5cm\epsfbox{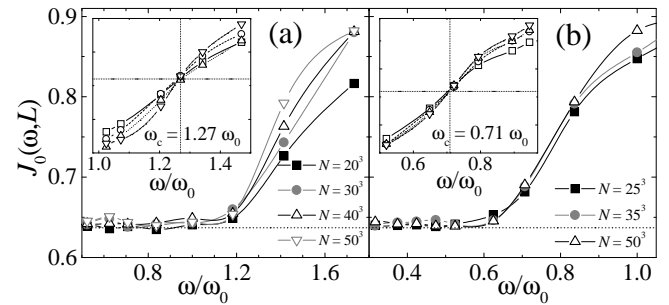} 
\parbox{8.5cm}{\caption[]{\small Level statistics results for
our model system with (a) $\km /\kma=10^{-2}$ and (b) $\km
/\kma=10^{-3}$.  $J_0(\omega,N)$ is shown versus 
$\omega/\omega_0$ for different system sizes $N$.  Values of $J_0
\approx J_0^{\rm Wigner} = 0.637$ indicate extended modes, while
$J_0 \approx J_0^{\rm Poisson} = 1$ indicates localized modes.  
In the insets, magnifications of the transition regions are shown,
which allow to determine the transition frequencies $\omega_{\rm
c}$ from the intersection of the curves.}
\label{fig:4}}\end{figure} 
 
Experimentally, it is not settled 
whether the frequency $\omega_{\rm {peak}}$ of the boson peak is
close to the localization delocalization frequency $\omega_{\rm
c}$ or well below.  Scattering experiments on vitreous silica  
have been discussed controversially.  One interpretation yields localized modes 
at the boson peak frequency $\omega_{\rm peak} \approx 6-9$ THz \cite{foret},  
while others indicate the presence of propagating modes up to $\omega_{\rm c} 
\approx 20$ THz \cite{benassi} well above the boson peak.  
Very recent scattering experiments on glassy glycerol yield extended modes up to 
at least $\omega_c \approx 2.5 \, \omega_{\rm{peak}}$ \cite{mascio}.
The results from \cite{benassi} and \cite{mascio} are in line with the 
results of our model system.

In summary we have considered an idealized disordered vibrational model system,
consisting of a simple cubic lattice of coupled harmonic oscillators with random
scalar spring constants between neighbors.  The spring constants were
chosen from a power-law distribution $P(k) \propto 1/k$ above a positive cut-off
value $\km$.  We have motivated the model by the presence of free volume in 
glassy materials.  Since the model describes (surprisingly) well several 
non-trivial features of the boson peak in glassy systems, we believe that 
despite its simplicity it may capture essential parts of the physics of this 
odd phenomenon -- in particular as no single fit parameter was involved.

We like to thank Prof. Michael Klinger for illuminating
discussions on the subject.

\end{multicols} 

\begin{thebibliography} {50}
\bibitem{footnote} These values may be in a realistic range for several 
glasses.  For example, we can approximate $a_{\rm min}=1.5$ \AA{} as a 
sensible Si-O closest distance and $a_{\rm max} \approx 5$ \AA{} 
as a reasonable maximum neighbor distance, e.g. across a sixfold ring in
SiO$_2$.  When we further assume a decay length $a^*$ around $0.5$
\AA, which does not seem unreasonable for the short-range potentials of
Si$O_2$ we can arrive at values for $\km/\kma = 
\exp[(a_{\rm{min}}-a_{\rm{max}})/a^*]$ around $10^{-3}$.
\end{thebibliography}
\end{document}